\documentclass{ws-procs975x65}

\newcommand*{\re}{\mathop{\mathrm{Re}}}
\newcommand*{\erfc}{\mathop{\mathrm{erfc}}}

\begin{document}

\title{Multifractality of the multiplicative autoregressive point processes}

\author{B. Kaulakys, M. Alaburda, V. Gontis and T. Meskauskas}

\address{Institute of Theoretical Physics and Astronomy of Vilnius University, \\ 
A. Gostauto 12, \\
LT-01108 Vilnius, Lithuania\\
E-mail: kaulakys@itpa.lt}  

\maketitle

\abstracts{
Multiplicative processes and multifractals have earned increased popularity in applications ranging from hydrodynamic 
turbulence to computer network traffic, from image processing to economics. 
We analyse the multifractality of the recently proposed point process models generating the signals exhibiting 
$1/f^{\beta}$ noise. 
The models may be used for modeling and analysis of stochastic processes in different systems.
We show that the multiplicative point process models generate multifractal signals, in contrast to the formally 
constructed signals with $1/f^{\beta}$ noise and signals consisting of sum of the uncorrelated components with a 
wide-range distribution of the relaxation times. 
}

\section{Introduction}

Multifractal models are used to account for scale invariance properties of various objects in different 
domains ranging from the energy dissipation in turbulent flows\cite{frisch:turbulence} to 
financial data\cite{bouchaud:pricing}. 
Healthy human heartbeat intervals exhibit multifractal properties rather than being fractal for a life-threatening 
condition, known as congestive heart failure\cite{ivanov:N399}.
Cerebral blood flow in healthy humans is also multifractal\cite{west:PA318}.

Scaling behavior has become a welcome careful description of complexity in many fields including natural phenomena, 
human heart rhythm in biology, spatial repartition of faults in geology, as well as human activities such as traffic 
in computer networks and financial markets. 
The multifractal formalism has received much attention as one of the most popular frameworks to describe and analyse 
signals and processes that exhibit scaling properties, covering and connecting both the local scaling and the global 
one in terms of sample moments.

 
The purpose of this paper is to analyse the multifractality of signals exhibiting $1/f^{\beta}$ noise generated by
different techniques and, especially, of the point processes with $1/f^{\beta}$ power spectral density\cite{KM98,KGA}.

First of all, however, we will analyse the multifractality of the signal constructed by the inverse fast Fourier
transform\cite{TK}. Using this method we can generate signals with any desirable slope $\beta$ of the power spectral
density $S(f)\sim 1/f^{\beta}$.

We calculate a generalized $q$th order height-height correlation function (GHCF) $F_{q}(t)$ defined as\cite{fractality}
\begin{equation}
F_{q}(t)=\langle|I(t^{\prime}+t)-I(t^{\prime})|^{q}\rangle^{1/q}, 
\label{ghcf}
\end{equation}
where the angular brackets denote the time average. The GHCF $F_{q}(t)$ characterizes the correlation properties of the 
signal $I(t)$, and for a multiaffine signal a power-law behavior like 
\begin{equation}
F_{q}(t)\sim t^{H_{q}}
\label{ghcf_approx}
\end{equation}
is expected. Here $H_{q}$ is the generalized $q$th order Hurst exponent. 
If $H_{q}$ is independent on $q$, a single scaling exponent $H_{q}$ is involved and the signal $I(t)$ is said to be 
monofractal\cite{fractality}. 
If $H_{q}$ depends on $q$, the signal is considered to be multifractal.

\begin{figure}[tbh]
\begin{center}
\includegraphics[width=.5\textwidth]{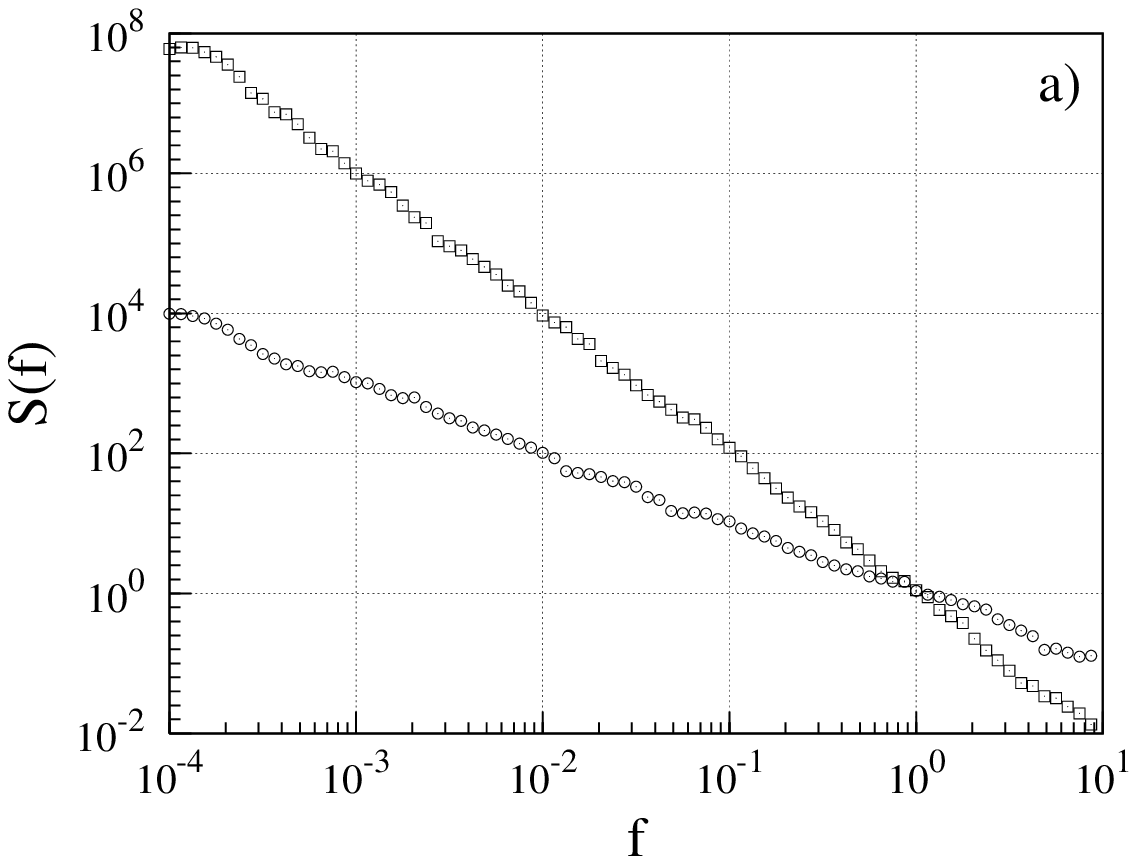}
\hspace{-10pt}
\includegraphics[width=.5\textwidth]{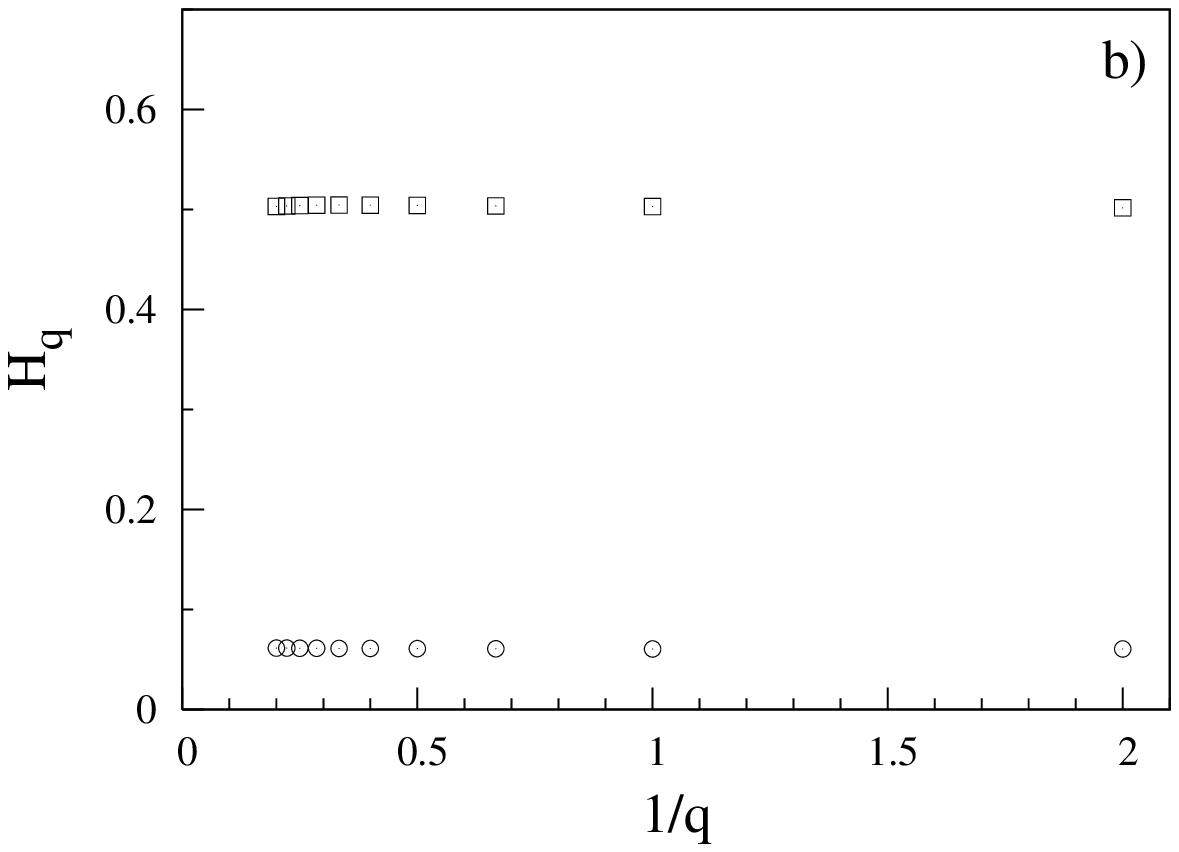}\\
\vspace{10pt}
\includegraphics[width=.5\textwidth]{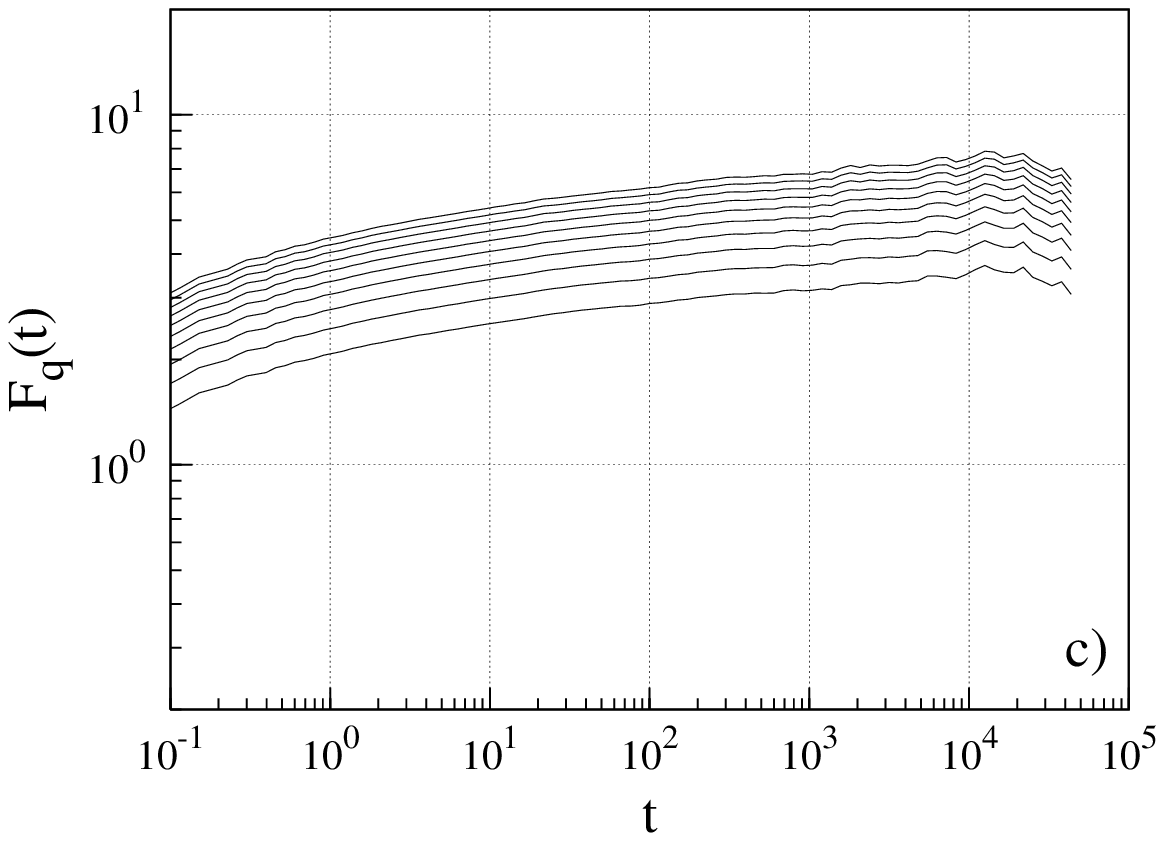}
\hspace{-10pt}
\includegraphics[width=.5\textwidth]{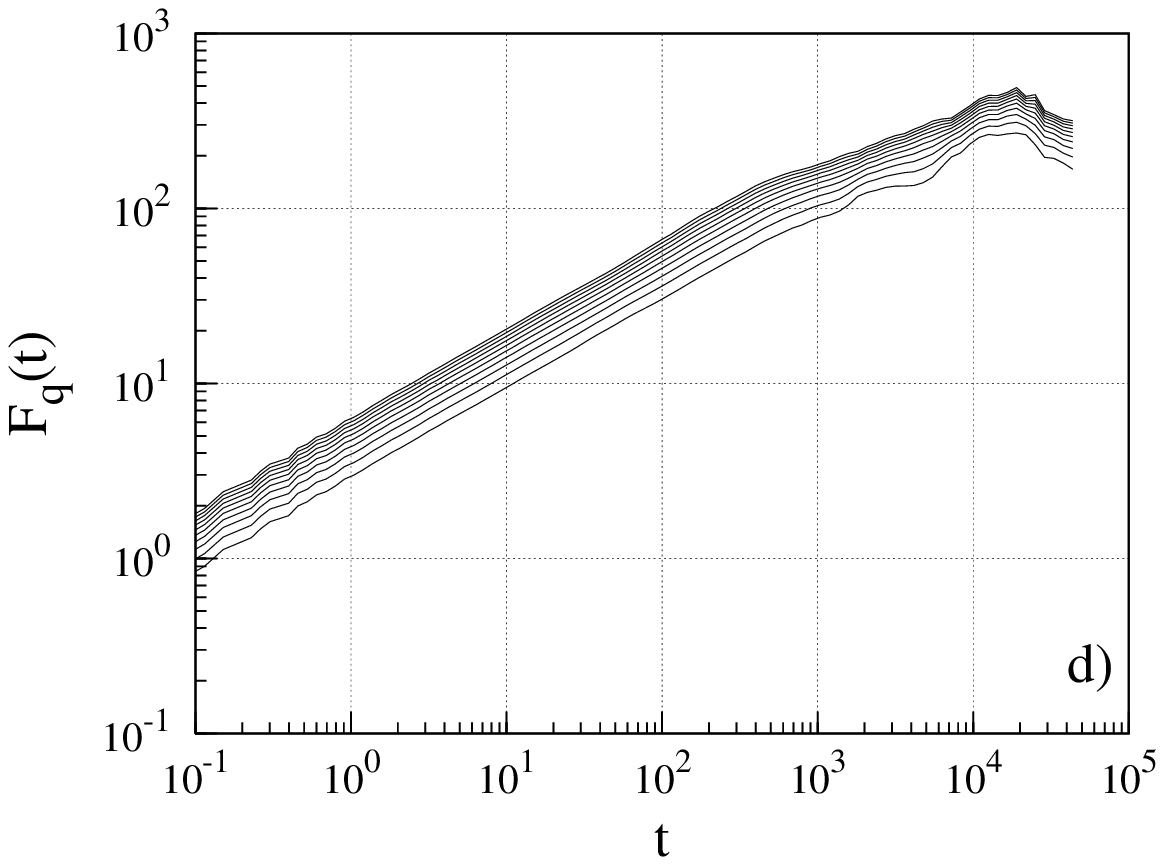}\\
\end{center}
\vspace{-10pt}
\caption{a) Power spectral density and b) the generalized Hurst exponents $H_{q}$ versus $1/q$ in the scaling regime 
$1<t<1000$ for the slopes $\beta=1$ (open circles) and $\beta=2$ (open squires). 
The signal of $10^{6}$ points was generated and averaged over $10$ realizations.
c) and d) show GHCF $F_{q}(t)$ versus time $t$ for the same parameters $\beta=1$ 
and $\beta=2$, respectively. }
\label{from_spectrum_ghcf}
\end{figure}

In figure \ref{from_spectrum_ghcf} a) we present a power spectral densities with the different slopes $\beta$ and in 
figure \ref{from_spectrum_ghcf} b) we show the Hurst exponents, calculated from GHCF using linear regression dependence 
on $1/q$ of the signals formally constructed by the inverse Fourier transform. 
In figure \ref{from_spectrum_ghcf} c) and d) corresponding GHCF $F_{q}(t)$ versus time $t$ are shown.
We see that Hurst exponent $H_{q}$ does not depend on $q$, which indicates that the signal is monofractal.

\section{Stochastic multiplicative point process}

In many cases the intensity of some signals or currents can be represented by a sequence of random
(however, as a rule, mutually correlated) pulses or elementary events $A_{k}(t-t_{k})$, 
\begin{equation}
I(t)=\sum_{k}A_{k}(t-t_{k}).  
\label{signal_pulse}
\end{equation}
Here the function $A_{k}(\phi)$ represents the shape of the $k$ pulse making an influence on the signal $I(t)$ 
in the region of the transit time $t_{k}$. 
We will be interested in the processes with the power-law distribution of the power spectral density at low frequencies.
It is easy to show that the shapes of the pulses mainly influence the high frequency, $f\gtrsim 1/\Delta t_{p}$, with 
$\Delta t_{p}$ being the characteristic pulse length, power spectral density, while fluctuations of the pulse 
amplitudes result, as a rule, in the white or the Lorentzian but not $1/f$ noise\cite{white_lorentz}. 
Therefore, we restrict our analysis to noise due to correlations between the transit times $t_{k}$. 
In such approach we can replace the function $A_{k}(t-t_{k})$ by the Dirac delta function 
and then express the signal as 
\begin{equation}
I(t)=\bar{a}\sum_{k}\delta(t-t_{k}),  
\label{signal}
\end{equation}
with $\bar{a}$ being an average contribution to the signal of one pulse. 
This model\cite{KM98} also corresponds to the flow of identical objects: electrons, photons, cars, and so on, 
and is called the point process model. 
Point processes arise in different fields, such as physics, economics, cosmology, ecology, neurology, seismology, 
traffic flow, signaling and telecom networks, and the Internet (see e.g., 
papers\cite{white_lorentz,point_process} and references herein). 

The power spectrum of the point process signal is described completely by the set of the interevent intervals 
$\tau_{k}=t_{k+1}-t_{k}$. 
Moreover, the low frequency noise is defined by the statistical properties of the signal at a large-time-scale, 
i.e., by the fluctuations of the time difference 
\begin{equation}
\Delta(k;q)\equiv t_{k+q}-t_{k}=\sum_{i=k}^{k+q-1}\tau_{i}
\label{delta_kq}
\end{equation}
at large $q$, determined by the slow dynamics of the average interpulse time $\tilde{\tau}_{k}(q) =\Delta(k;q)/q$ between 
the occurrence of pulses $k$ and $k+q$. 
Quite generally the dependence of the average interevent time $\tilde{\tau}_{k}$ may be described by the general 
Langevin equation. The Langevin equation may be written down in the actual time $t$ or, equivalently, in the space of 
the occurrence numbers $k$ with the drift coefficient $h(\tilde{\tau}_{k})$ and a multiplicative noise 
$g(\tilde{\tau}_{k})\xi(k)$,
\begin{equation}
\frac{d\tilde{\tau}_{k}}{dk}=h(\tilde{\tau}_{k})+g(\tilde{\tau}_{k})\xi(k).  
\label{Langevin}
\end{equation}
Here we interpret $k$ as a continuous variable while the white Gaussian noise $\xi(k)$ satisfies the standard condition 
\begin{equation}
\langle\xi(k)\xi(k^{\prime})\rangle=\delta(k-k^{\prime })
\label{white_correlation}
\end{equation}
with the brackets $\langle\ldots\rangle$ denoting the averaging over the realizations of the process. 
We understand the equation \eqref{Langevin} in It\^{o} interpretation.

Transition from the occurrence numbers $k$ to the actual time $t$ in Eq.~\eqref{Langevin} may be fulfilled using the 
relation $dt=\tilde{\tau}_{k}dk$\cite{KR}.

The particular sequence of the interevent times $\tau_{k}$ may be superimposed by some additional noise or stochasticity,
e.g., $\tau_{k}$ may be determined by the Poisson distribution
\begin{equation}
P(\tau_{k})=\frac{1}{\tilde{\tau}_{k}}e^{-\tau_{k}/\tilde{\tau}_{k}}
\label{poisson_distribution}
\end{equation}
with the slowly, according to Eq.~\eqref{Langevin}, changeable average interevent time $\tilde{\tau}_{k}$. Such additional
stochasticity do not influence the long-range statistical properties and the low frequency spectra of the process.
Therefore, further we will restrict the analysis to the processes generated by Eq.~\eqref{Langevin} and will identify
$\tau_{k}$ with $\tilde{\tau}_{k}$.

\subsection{Power spectral density}

The point process is entirely defined by the occurrence times $t_{k}$. 
The power spectral density of the point process \eqref{signal} may be expressed as 
\begin{equation}
S(f)=\lim_{T\to\infty}\left\langle\frac{2}{T}\left|\int\limits_{t_{i}}^{t_{f}}I(t)e^{-i2\pi ft}dt\right|^{2}\right\rangle
=\lim_{T\to\infty}\left\langle\frac{2\bar{a}^{2}}{T}\sum_{k}\sum_{q=k_{\min}-k}^{k_{\max}-k}e^{i2\pi f\Delta(k;q)}
\right\rangle,
\label{spectrum}
\end{equation}
where $t_{i}$ and $t_{f}$ are initial and final observation times, $T=t_{f}-t_{i}\gg\omega^{-1}$ is the whole observation 
time and $\omega=2\pi f$.
Here $k_{\min}$ and $k_{\max}$ are minimal and maximal values of index $k$ in the interval of observation $T$ and 
the brackets $\langle\ldots\rangle$ denote the averaging over realizations of the process. 

For the interpulse intervals described by the Langevin equation \eqref{Langevin} we use a perturbative solution in the
vicinity of $\tau_{k}$. After replacing the averaging over $k$ by the averaging over the distribution $P_{k}(\tau_{k})$ 
of the interpulse times $\tau_{k}$, we have the power spectrum\cite{KGA}
\begin{equation}
S(f)=2\bar{I}^{2}\frac{\bar{\tau}}{\sqrt{\pi}f}\int\limits_{0}^{\infty}P_{k}(\tau_{k})\re[{e}^{-i(x-\frac{\pi}{4})}
\erfc\sqrt{-ix}]\frac{\sqrt{x}}{\tau_{k}}d\tau_{k},  
\label{Langevin_spectrum}
\end{equation} 
where $\bar{I}$ and $\bar{\tau}$ are the averages of the signal and the interpulse times, respectively, and 
$x=\pi f\tau_{k}^{2}/h(\tau_{k})$. 

The replacement of the averaging over $k$ and over realizations of the process by the averaging over the distribution 
of the interpulse times $\tau_{k}$, $P_{k}\left( \tau_{k}\right)$, is possible when the process is ergodic. 
Ergodicity is usually a common feature of the stationary process described by the general Langevin equation.

According to Eq.~\eqref{Langevin_spectrum} the small interpulse times and the clustering of the pulses make the greatest 
contribution to $1/f^{\beta }$ noise. 
The power-law spectral density is very often related with the power-law behavior of other characteristics of the signal, 
such as autocorrelation function, probability densities and other statistics, and with the fractality of the signals, 
in general\cite{power-law}. 
Therefore, we investigate the power-law 
dependences of the drift coefficient and of the distribution density on the
time $\tau _{k}$ in some interval of the small interpulse times, i.e., 
\begin{equation}
h(\tau_{k})=\gamma\tau_{k}^{\delta},\quad P_{k}(\tau_{k})=C\tau_{k}^{\alpha},\quad\tau_{\min}\leq\tau_{k}\leq\tau_{\max },
\label{power}
\end{equation}
where the coefficient $\gamma $ represents the rate of the signal's nonlinear relaxation
and $C$ has to be defined from the normalization. 

The simplest and the well-known process generating the power-law probability distribution function for $\tau_{k}$ 
is a multiplicative stochastic process with $g(\tau_{k})=\sigma\tau_{k}^{\mu}$ and $\delta=2\mu-1$, written 
as\cite{KGA,GK} 
\begin{equation}
\tau_{k+1}=\tau_{k}+\gamma\tau_{k}^{2\mu-1}+\sigma\tau_{k}^{\mu}\varepsilon_{k}.  
\label{multiplicative}
\end{equation}
Here $\gamma$ represents the nonlinear relaxation of the signal, while $\tau_{k}$ fluctuates due to the perturbation 
by normally distributed uncorrelated random variables $\varepsilon_{k}$ with a zero expectation and unit variance
and $\sigma$ is a standard deviation of the white noise.

Eq.~\eqref{multiplicative} is the difference (discrete) version of the differential equation \eqref{Langevin}.
On the other hand, it is the generalization of the simple autoregressive model of $1/f$ noise\cite{KM98} (see also
Eq.~\eqref{additive}) and represents quite general evolution of the interevent time with the nonlinear drift 
$h(\tau_{k})=\gamma\tau_{k}^{2\mu-1}$ and the multiplicative noise $\sigma\tau_{k}^{\mu}\varepsilon_{k}$, resulting 
in the $1/f^{\beta}$ noise and power-law distribution \eqref{power} of the interevent time $\tau_{k}$ with the exponent
$\alpha=2\gamma/\sigma^{2}-2\mu$. Indeed, the power spectrum for the process \eqref{multiplicative}, when 
$\gamma/(\pi\tau_{\max}^{2-\delta})\ll f\ll \gamma/(\pi\tau_{\min}^{2-\delta})$, is\cite{KGA} 
\begin{equation}
S(f)=\frac{(2+\alpha)(\beta-1)\bar{a}^{2}\Gamma(\beta-1/2)}{\sqrt{\pi}\alpha(\tau_{\max}^{2+\alpha}-
\tau_{\min}^{2+\alpha})\sin(\pi\beta/2)}\left(\frac{\gamma}{\pi}\right)^{\beta-1}\frac{1}{f^{\beta}}, 
\label{multiplicative_spectrum}
\end{equation}
where 
\begin{equation} 
\alpha=\frac{2\gamma}{\sigma^{2}}-2\mu,\quad\beta=1+\frac{\alpha}{3-2\mu},\quad\frac{1}{2}<\beta<2.   
\label{beta}
\end{equation}
For $\mu =1$ we have a completely multiplicative point process when the stochastic change of the interpulse time is 
proportional to itself. Another case of interest concerns $\mu=1/2$, then we have the Brownian motion of the interevent 
time with the linear relaxation of the signal $I\simeq\bar{a}/\tau$.

\begin{figure}[tbh]
\begin{center}
\includegraphics[width=.5\textwidth]{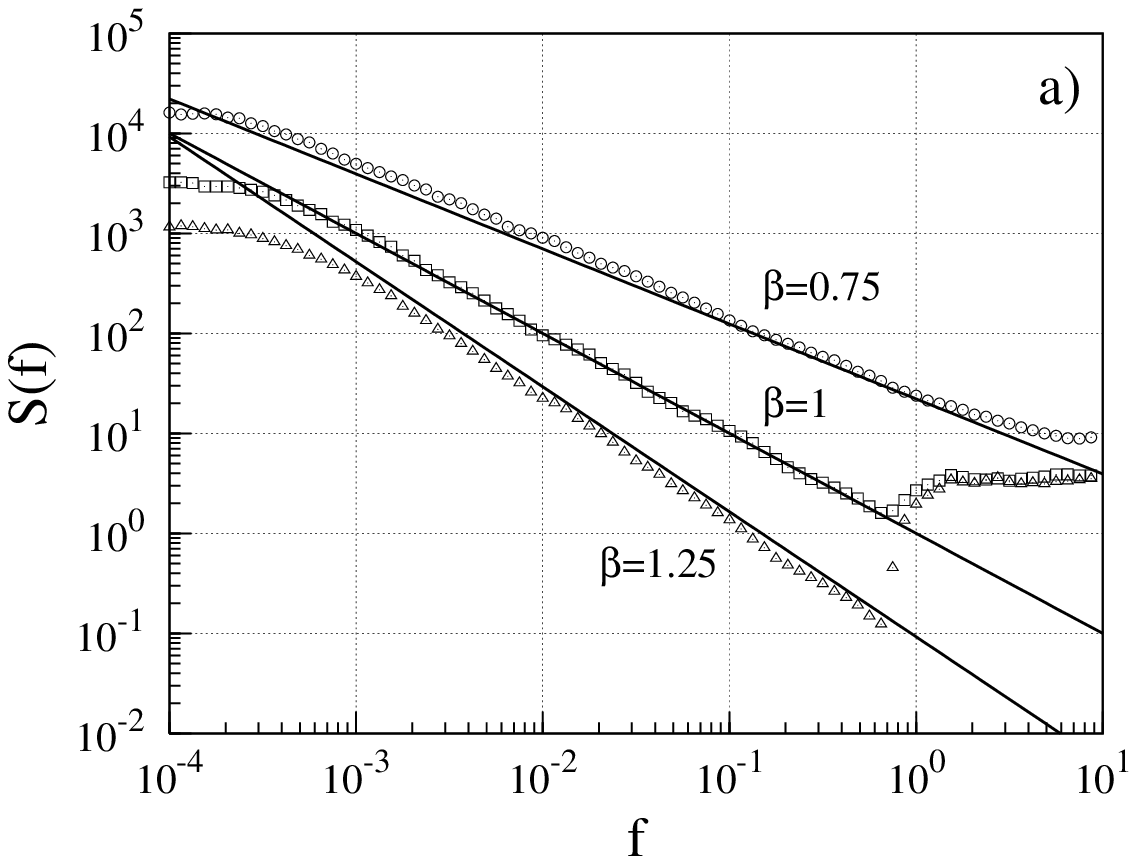}
\hspace{-10pt}
\includegraphics[width=.5\textwidth]{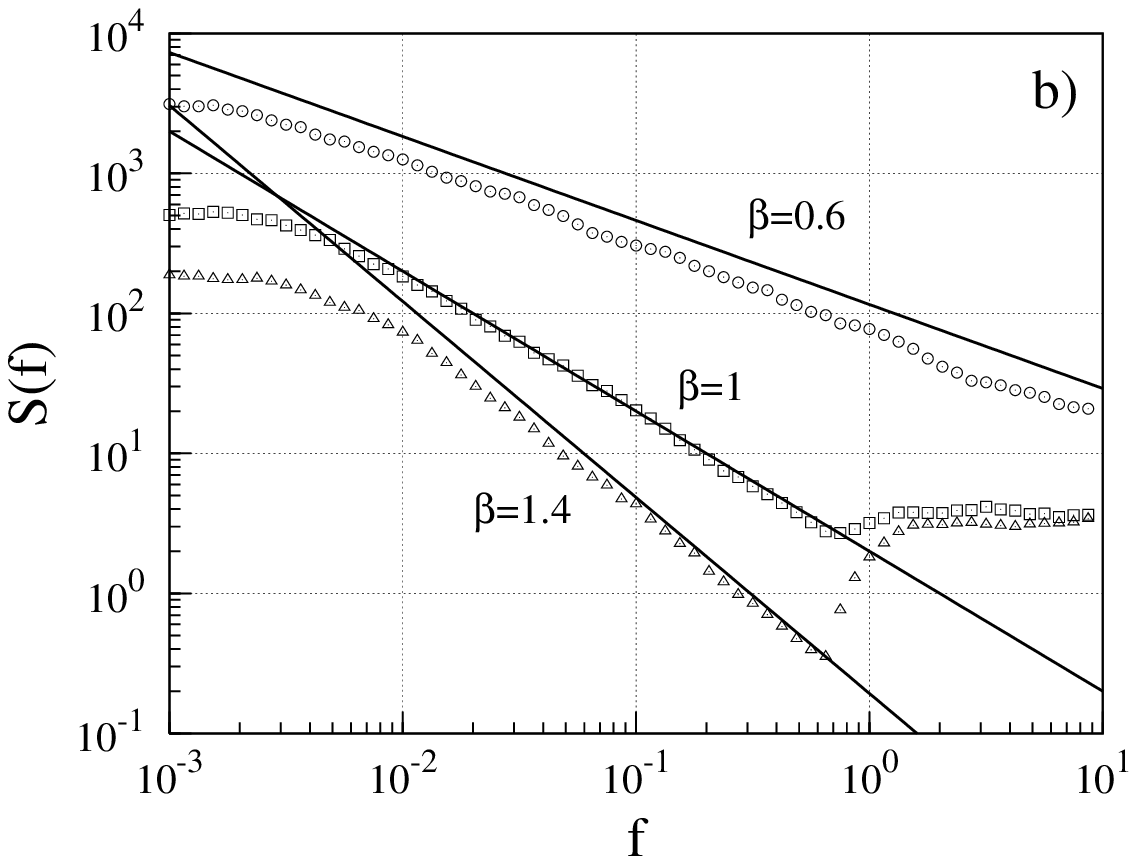}
\end{center}
\vspace{-10pt}
\caption{Power spectral density \eqref{spectrum} vs frequency of the signal generated
by Eqs.~\eqref{signal} and \eqref{multiplicative} with the parameters a) $\mu=0.5$, 
$\sigma=0.02$ and different relaxations of the
signal $\gamma=0.0001$ (open circles), $0.0002$ (open squares) and $0.0003$ (open triangles); and b) $\mu=1$, 
$\sigma=0.1$ and $\gamma=0.008$ (open circles), $0.01$ (open squares) and $0.012$ (open triangles). 
We restrict the diffusion of the interevent time
in the interval $\tau_{\min}=10^{-6}$, $\tau_{\max}=1$
with the reflective boundary condition at $\tau_{\min}$ and
transition to the white noise, $\tau_{k+1}=\tau_{\max}+\sigma\varepsilon_{k}$, 
for $\tau_{k}>\tau_{\max}$ and $100$ realizations with $10^{6}$ $t_{k}$ points each were used.
The solid lines represent the analytical results according to Eq.~\eqref{multiplicative_spectrum}.}
\label{point_process_spectrum}
\end{figure}

Figure \ref{point_process_spectrum} represents the spectral densities \eqref{spectrum} with different slopes 
$\beta$ of the signals generated numerically according to Eqs. \eqref{signal} and \eqref{multiplicative} for 
different parameters of the model. 
We see that the simple iterative equation \eqref{multiplicative} with the multiplicative noise produces the signals 
with the power spectral density of different slopes, depending on the parameters of the model. 
The agreement of the numerical results with the approximate theory is quite good.

\subsection{Multifractal point processes}

The multifractal formalism has received much attention recently as one of the most popular frameworks 
to describe and analyse signals and processes that exhibit the scaling properties.

Fractality of the point process can be investigated by transition from the point process to the stochastic signal 
$I(t)$, using the rectangular constant area pulses, instead of the Dirac delta functions. 
The stochastic signal will have the same fractal properties as the origin point process. 

\begin{figure}[tbh]
\begin{center}
\includegraphics[width=.5\textwidth]{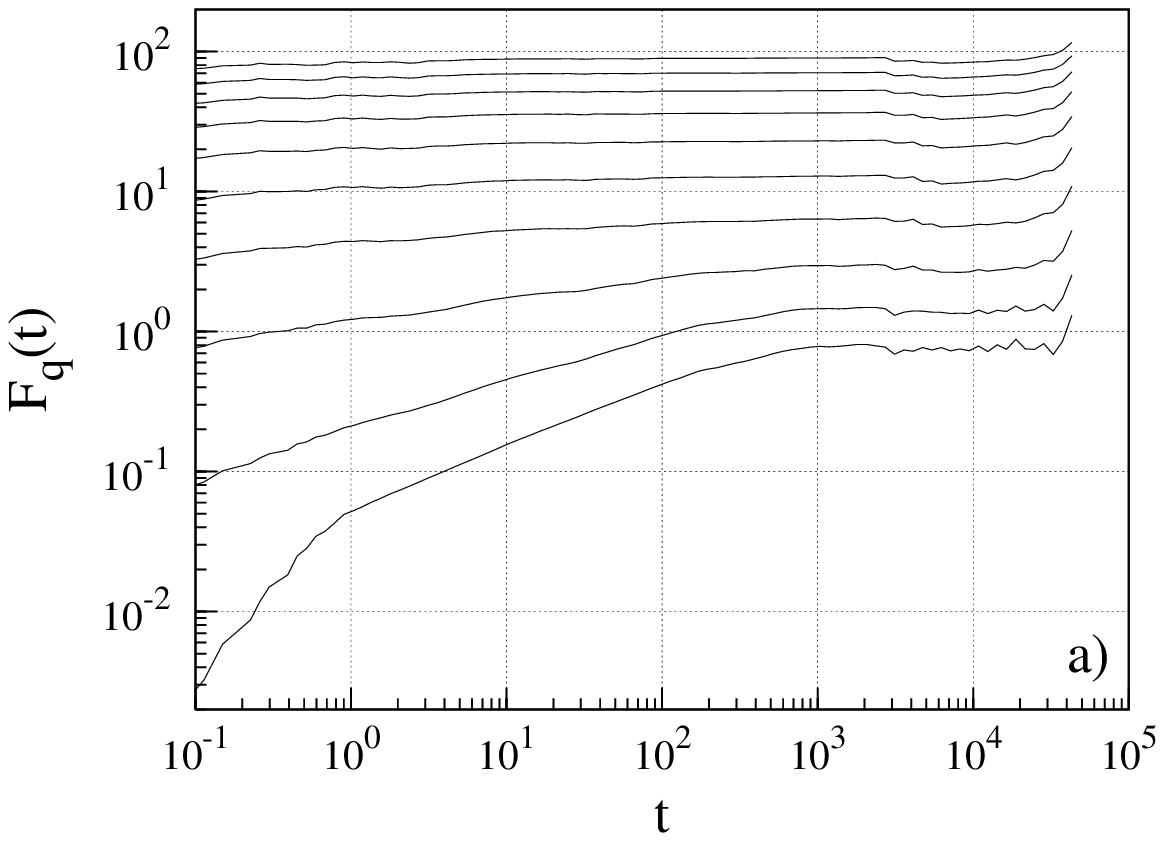}
\hspace{-10pt}
\includegraphics[width=.5\textwidth]{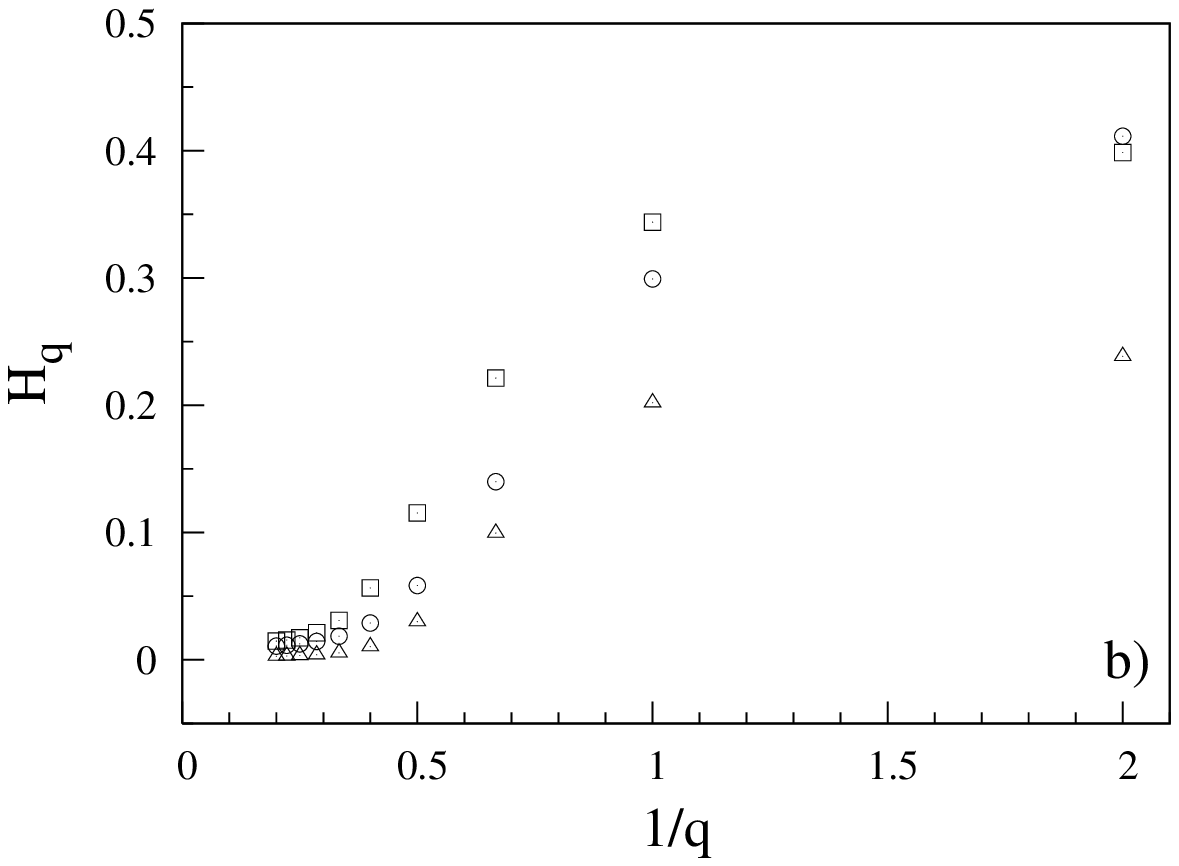}
\end{center}
\vspace{-10pt}
\caption{a) Generalized height-height correlation function $F_{q}(t)$ versus time $t$ for the signal \eqref{signal}
and \eqref{multiplicative} with $\mu=0.5$, $\bar{\tau}=1$, $\sigma=0.02$, $\gamma=0.0002$, and $q=0.5,1,1.5,\ldots,5$ 
from bottom to top. 
b) The generalized Hurst exponents $H_{q}$ versus $1/q$ in the scaling regime $1<t<1000$ for the $\bar{\tau}=1$ and 
$\mu=0.5$, $\sigma=0.02$, $\gamma=0.0002$ (open circles); $\mu=0.5$, $\sigma=0.02$, $\gamma=0.0003$ (open squares); 
$\mu=1$, $\sigma=0.1$, $\gamma=0.008$ (open triangles).}
\label{point_process_ghcf}
\end{figure}

In figure \ref{point_process_ghcf} a) we present the GHCF as a function of the time interval $t$, and in figure 
\ref{point_process_ghcf} b) we show the Hurst exponents calculated from GHCF using the linear regression dependence on 
$1/q$ for different parameters $\mu$, $\sigma$, and $\gamma$. 
We observe the clear multifractal behavior since the slopes of the log-log plot of GHCF are depending on $q$. 

Another interesting case is a sequence of transit times with random increments of the time intervals between pulses, 
$\tau_{k}=\tau_{k-1}+\sigma\varepsilon_{k}$. 
It is natural to restrict in some way the infinite Brownian increase or decrease of the interpulse times $\tau_{k}$, 
e.g., by the introduction of the relaxation to the average interpulse time $\bar{\tau}$ rate $\gamma$. 
So, we have an additive point process 
\begin{equation}
\tau_{k}=\tau_{k-1}-\gamma(\tau_{k-1}-\bar{\tau})+\sigma\varepsilon_{k}. 
\label{additive}
\end{equation}
This model generates the process with $1/f$ noise\cite{KM98}, and may be useful for modeling and analysing different 
systems (see references in paper\cite{KGA}).
Introduction of the reflective boundary condition at $\tau_{\min}>0$ avoids the formation of clusters and leads to 
$1/f^{2}$ noise. 

\begin{figure}[tbh]
\begin{center}
\includegraphics[width=.5\textwidth]{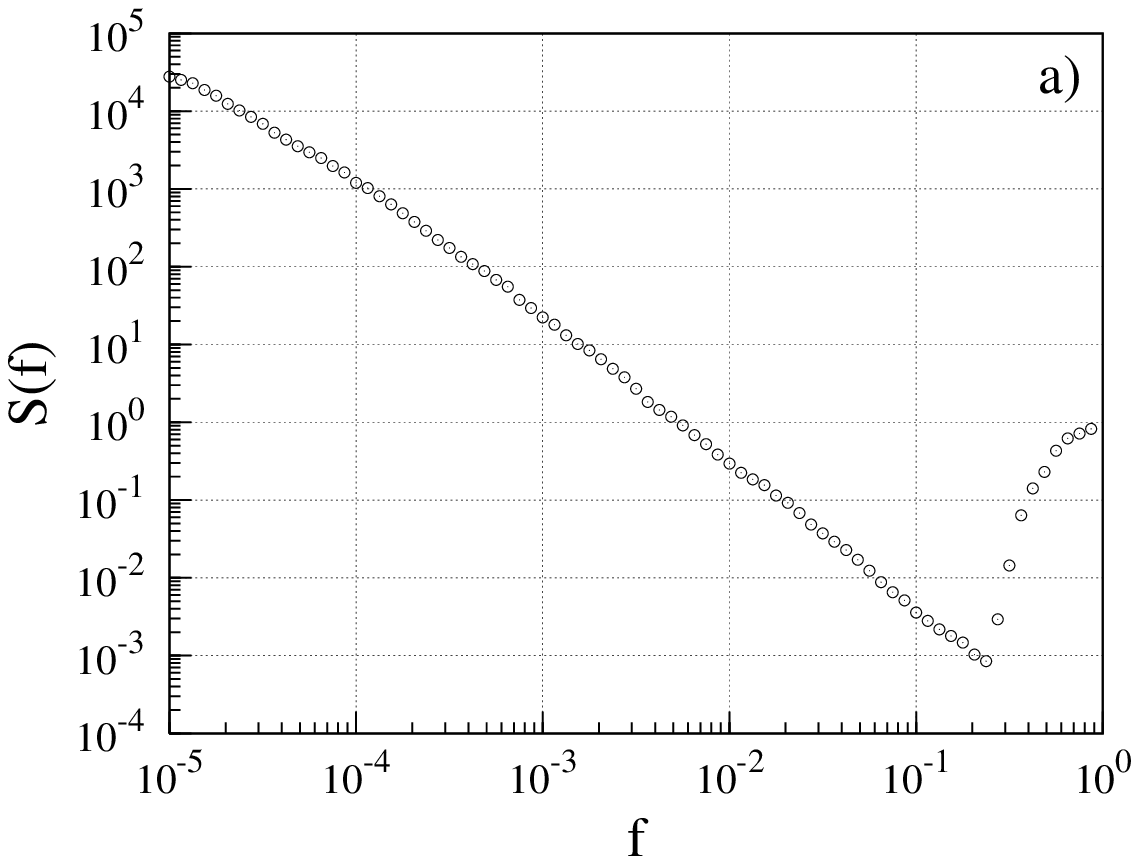}
\hspace{-10pt}
\includegraphics[width=.5\textwidth]{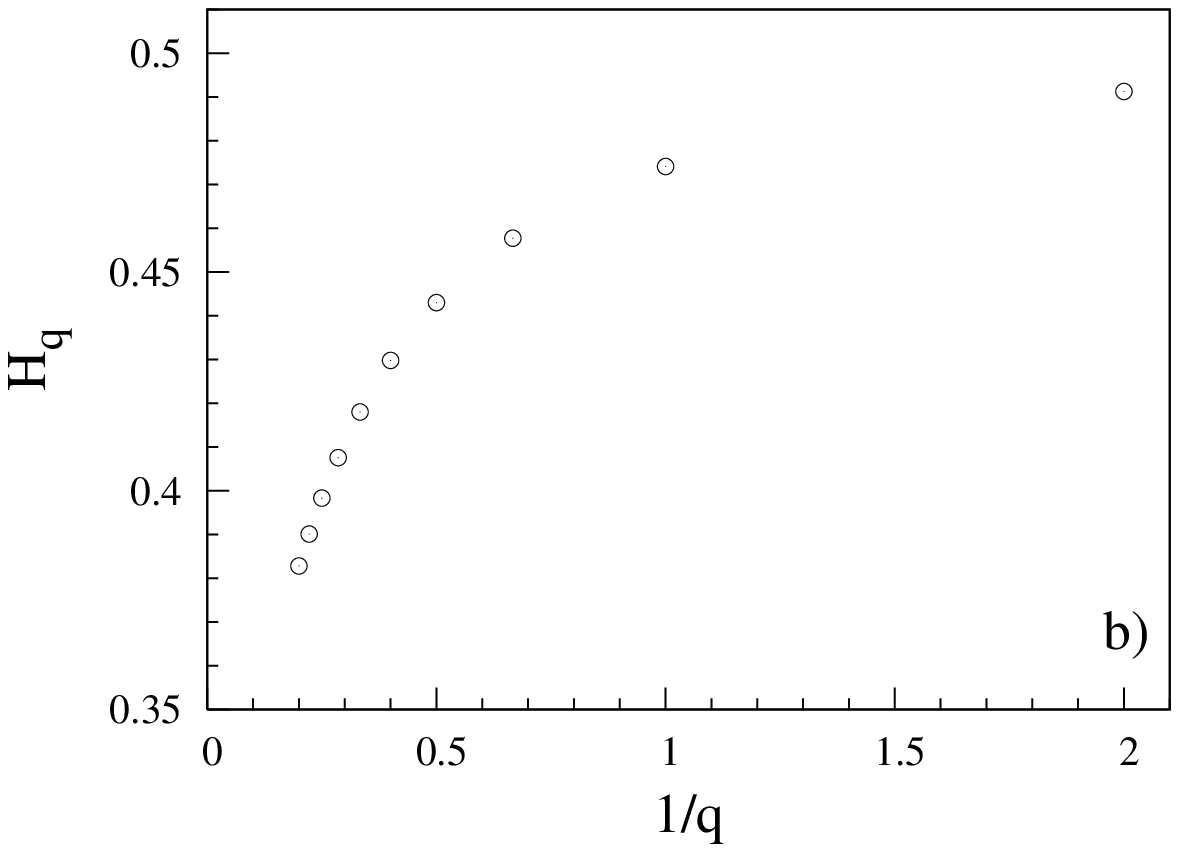}
\end{center}
\vspace{-10pt}
\caption{a) Power spectral density and b) the generalized Hurst exponents of the additive point process \eqref{additive} 
with the reflecting boundary condition at $\tau_{\min}=0.1$ in the scaling regime $1<t<1000$. 
The signal of $10^{6}$ points was generated with the parameters $\bar{\tau}=1$, $\sigma=0.001$, and 
$\gamma=0.000001$.}
\label{point_1f2_ghcf}
\end{figure}

In figure \ref{point_1f2_ghcf} a) we present a power spectral density of the additive point process \eqref{additive} 
with the reflecting boundary condition and in figure \ref{point_1f2_ghcf} b) we show the generalized Hurst exponents. 
We observe the multifractal behavior of the additive point process.

\subsection{Monofractality of the white and the Gaussian noises}

It is well-known\cite{white_lorentz} that the point processes with the interevent time $\tau_{k}$ distributed according 
to Poisson distribution 
\begin{equation}
P(\tau_{k})=\frac{1}{\bar{\tau}}e^{-\tau_{k}/\bar{\tau}}
\label{poisson}
\end{equation}
generate the white noise, $S(f)=const$.

\begin{figure}[tbh]
\begin{center}
\includegraphics[width=.5\textwidth]{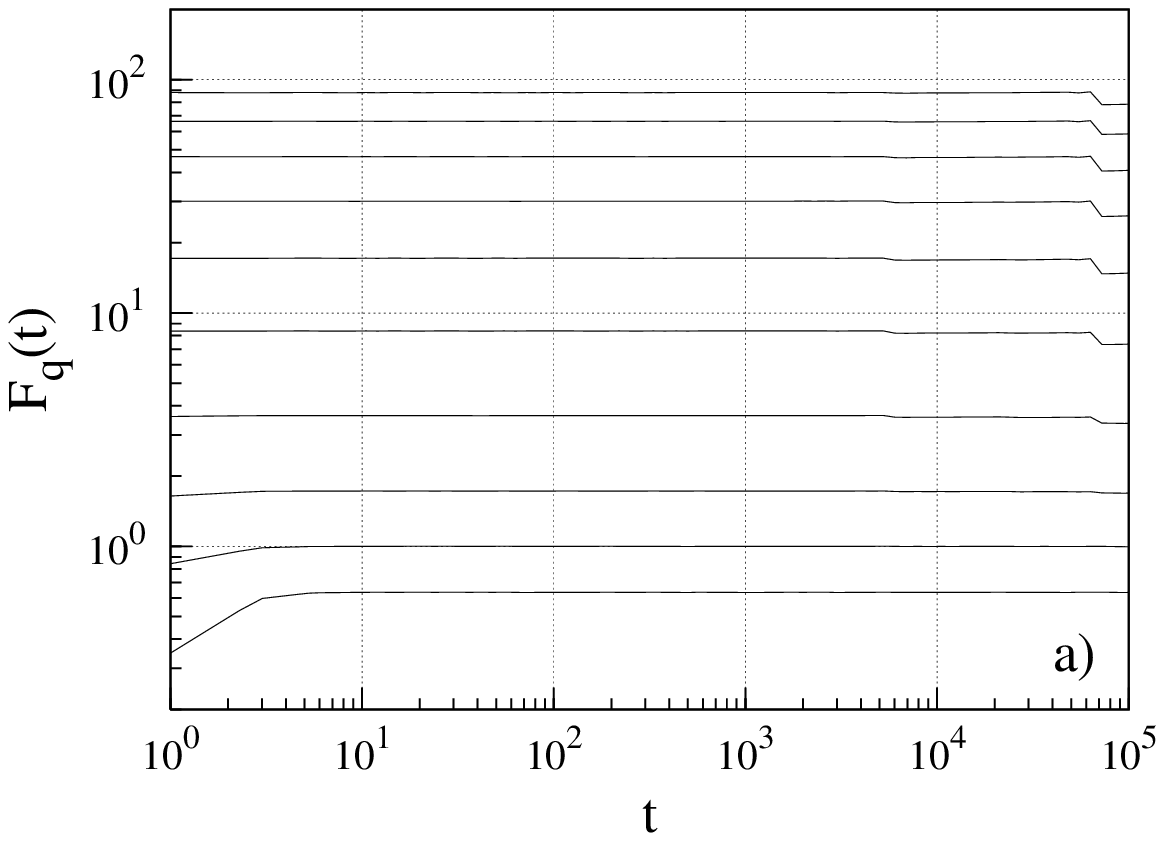}
\hspace{-10pt}
\includegraphics[width=.5\textwidth]{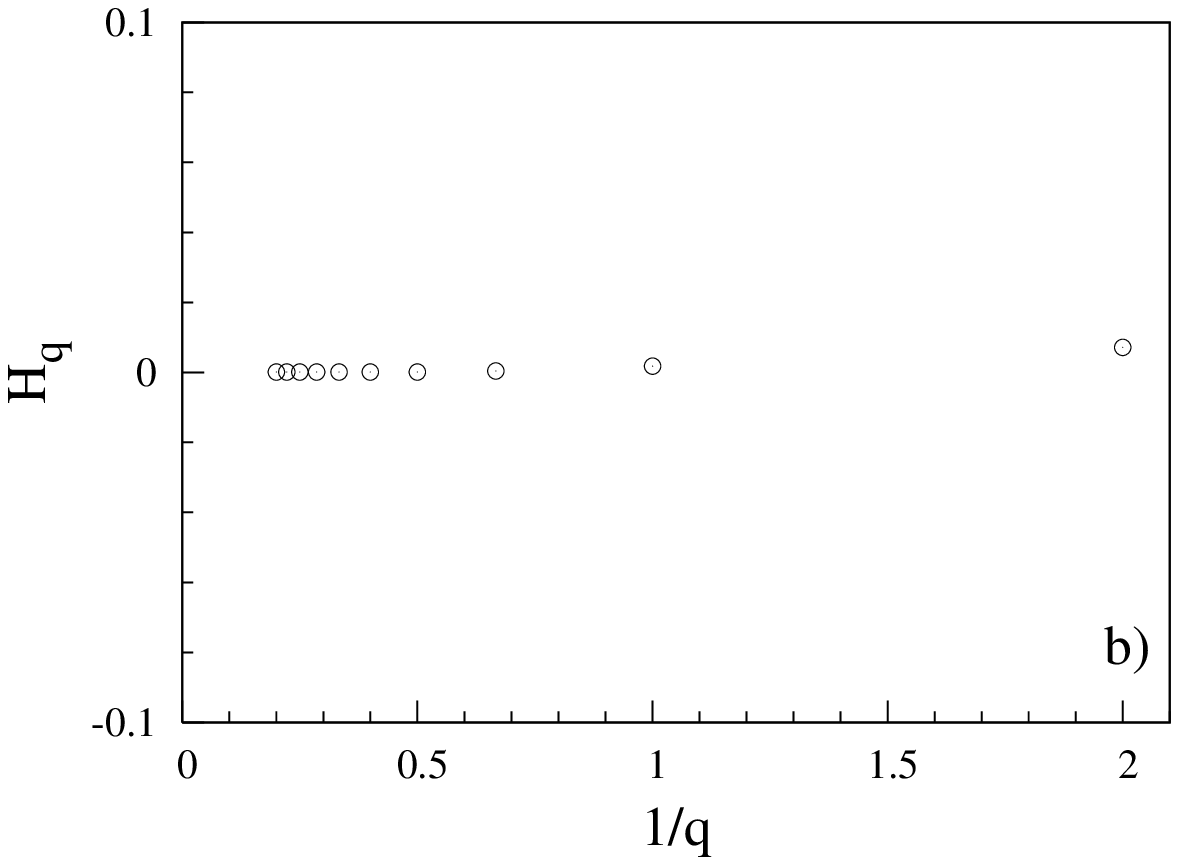}
\end{center}
\vspace{-10pt}
\caption{a) Generalized height-height correlation function $F_{q}(t)$ versus time $t$, and b) the generalized Hurst 
exponents $H_{q}$ versus $1/q$ of the point process with interevent time $\tau_{k}$ distributed according to Poisson 
distribution \eqref{poisson} in the scaling regime $1<t<1000$. 
The signal of $10^{6}$ points with the parameter $\bar{\tau}=1$ has been generated.}
\label{white_ghcf}
\end{figure}

In figure \ref{white_ghcf} a) we present the GHCF as a function of the time interval $t$ and in figure \ref{white_ghcf} 
b) we show the Hurst exponents for the white noise.
The Hurst exponents of the white noise are equal to zero. 
This demonstrates the absence of the scaling and that there is no correlation in time.

Noise with the power-law spectral density $1/f$ is often modeled as the sum of the Lorentzian spectra with 
the appropriate weights of a wide range distribution of the relaxation times $\tau^{rel}$. 
The signal may be expressed as a sum of $N$ uncorrelated components,
\begin{equation}
I(t)=\sum_{l=1}^{N}I_{l}(t)=\int\limits_{\gamma_{\min}}^{\gamma_{\max}}I(t,\gamma)g(\gamma)d\gamma,  
\label{lorentzian_signal}
\end{equation}
where $g(\gamma)$ is the distribution of the relaxation rates $\gamma=1/\tau^{rel}$, and every component $I_{l}$ 
satisfies the stochastic differential equation 
\begin{equation}
\dot{I}_{l}=-\gamma _{l}(I_{l}-\bar{I}_{l})+\sigma _{l}\xi _{l}(t).
\label{lorentzian_component}
\end{equation}
Here $\bar{I}_{l}$ is the average value of the signal component $I_{l}$, $\xi_{l}(t)$ is the $\delta$-correlated white 
noise, $\langle\xi_{l}(t)\xi_{l^{\prime}}(t^{\prime})\rangle=\delta_{l,l^{\prime}}\delta(t-t^{\prime})$, 
and $\sigma_{l}$ is the intensity (standard deviation) of the white noise. 
The steady-state solution of the stationary Fokker-Planck equation corresponding to Eq.~\eqref{lorentzian_component} 
for each component $I_{l}$ and the resulting signal \eqref{lorentzian_signal} yields the Gaussian distribution densities, 
however, the power spectrum may be of the power-law form when $\sigma^{2}(\gamma)g(\gamma)$ is constant or a power-law 
function\cite{KGA}.

\begin{figure}[tbh]
\begin{center}
\includegraphics[width=.5\textwidth]{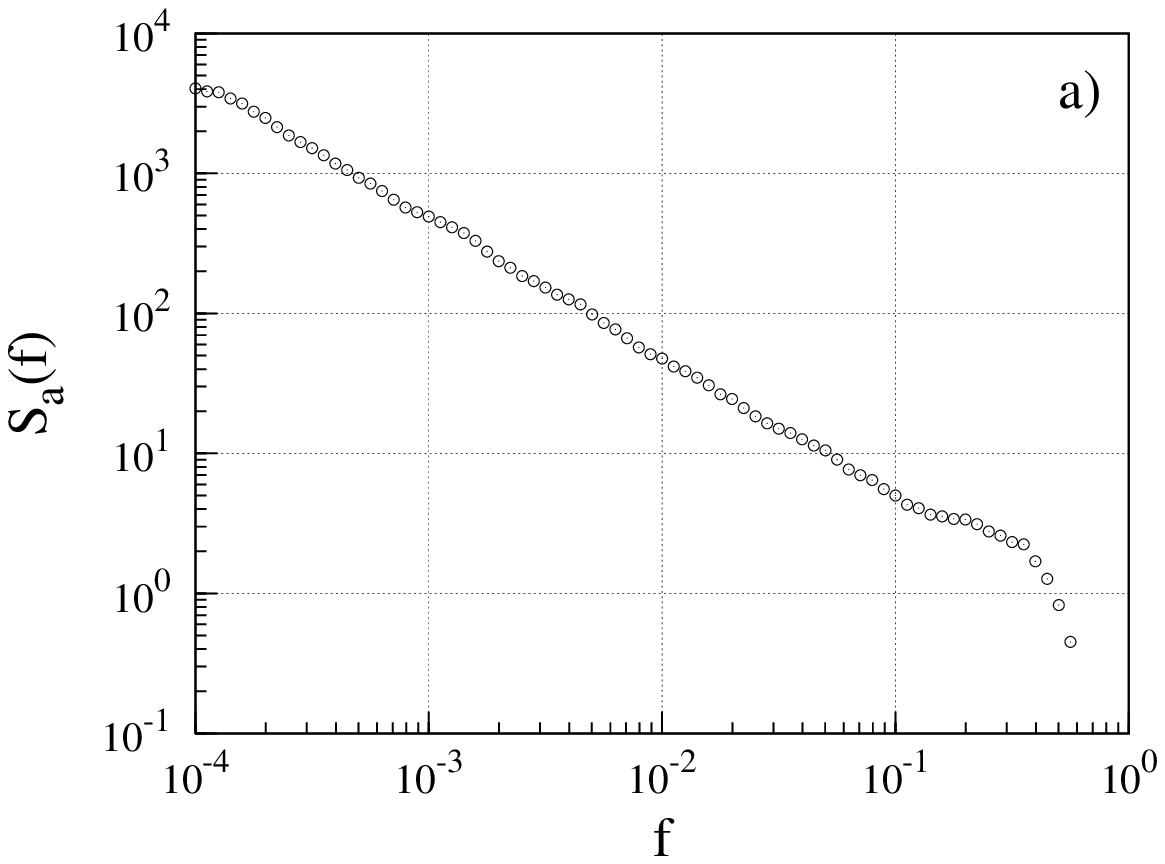}
\hspace{-10pt}
\includegraphics[width=.5\textwidth]{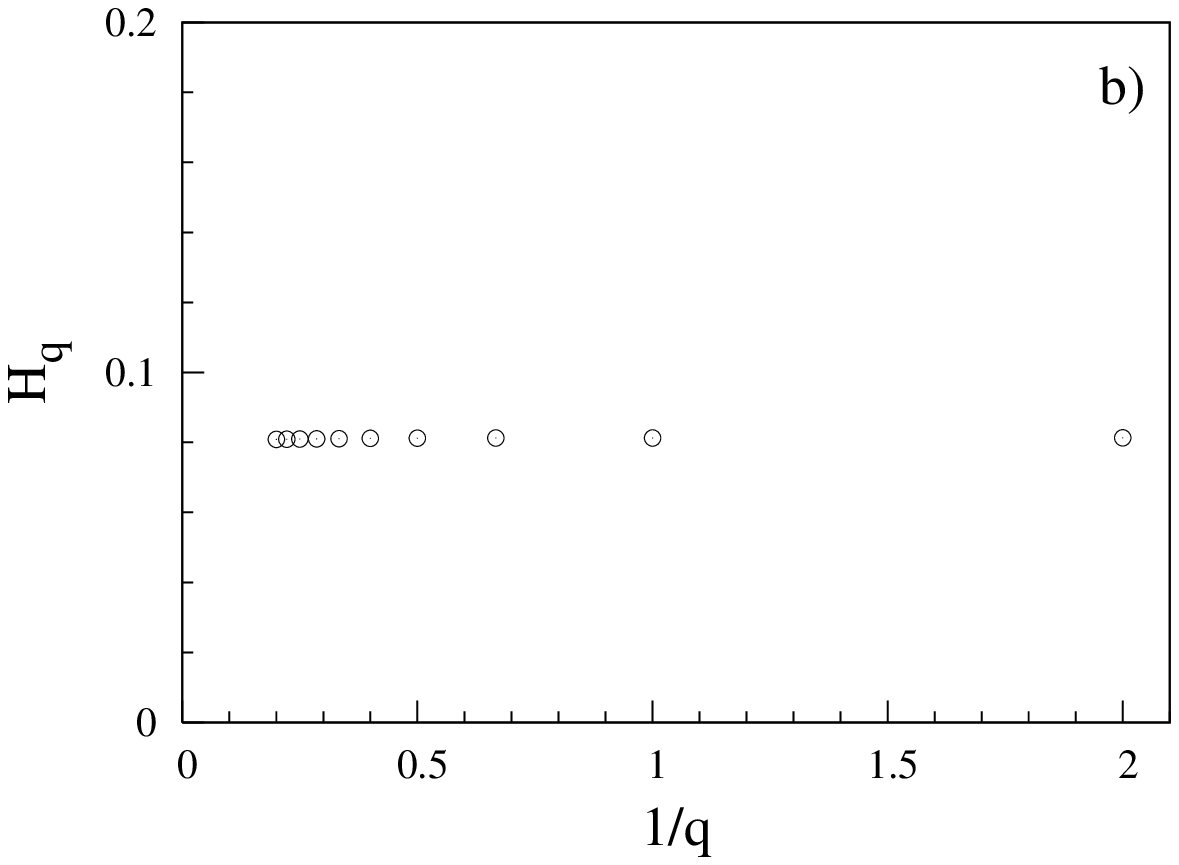}
\end{center}
\vspace{-10pt}
\caption{a) Power spectral density and b) the generalized Hurst exponents 
$H_{q}$ versus $1/q$ in the scaling regime $1<t<1000$. 
The signal of $10^{6}$ points was generated from $10$ components with the parameters $\bar{I}=20$, 
$\sigma^{2}(\gamma)g(\gamma)=10$, and uniform distribution of $\lg\gamma$ with $\gamma$ values in the interval 
$10^{-4}-1$.}
\label{lorentzian_ghcf}
\end{figure}

In figure \ref{lorentzian_ghcf} a) we present a power spectral density of the sum of the signals with a wide range 
distribution of the relaxation times $\tau^{rel}$ and in figure \ref{lorentzian_ghcf} b) we show the Hurst exponents, 
calculated from GHCF using linear regression dependence on $1/q$.
In the figures we observe $1/f$ behavior of the signal noise and clearly see that Hurst exponent $H_{q}$ does not 
depend on $q$, which shows that the signal \eqref{lorentzian_signal} is monofractal.

\section{Conclusions}

The multiplicative \eqref{multiplicative} and additive \eqref{additive} stochastic point processes may generate time 
series exhibiting the power spectral density $S(f)\sim 1/f^{\beta}$ and show clear multifractal behavior, however, 
the formally constructed by the inverse Fourier transform signal, Poisson white noise \eqref{poisson} and Gaussian, 
Eqs. \eqref{lorentzian_signal} and \eqref{lorentzian_component}, signal with the power spectral density $S(f)\sim 1/f$ 
are monofractal.

Therefore, the proposed\cite{KM98} and generalized\cite{KGA} point process models of $1/f^{\beta}$ noise may be used for
modeling of stochastic multifractal processes in different systems.

\section*{Acknowledgments}

We acknowledge support by the Lithuanian State Science and Studies Foundation and EU COST Action P10 ``Physics of
Risk''.


\begin{thebibliography}{0}
\bibitem{frisch:turbulence}
U.~Frisch, {\em Turbulence: The Legacy of A. N. Kolmogorov\/}, Cambridge
  University Press (1995).

\bibitem{bouchaud:pricing}
J.-P. Bouchaud and M.~Potters, {\em Theory of Financial Risk and Derivative
  Pricing\/}, Cambridge University Press (1999).

\bibitem{ivanov:N399}
P.~C. Ivanov, M.~G. Rosenblum, L.~A.~N. Amaral, Z.~R. Struzik, S.~Havlin, A.~L.
  Goldberger, and H.~E. Stanley, Nature {\bf 399}, 461 (1999);
P.~C. Ivanov, L.~A.~N. Amaral, A.~L. Goldberger et al., Chaos {\bf 11}, 641 (2001).

\bibitem{west:PA318}
B.~J. West, M.~Latka, M.~Glaubic-Latka, and D.~Latka, Physica A {\bf 318}, 453
  (2003).

\bibitem{KM98}  
B.~Kaulakys and T.~Me\v{s}kauskas, Phys. Rev. E \textbf{58}, 7013 (1998); 
B.~Kaulakys, Phys. Lett. A \textbf{257}, 37 (1999). 

\bibitem{KGA} B.~Kaulakys, V.~Gontis, and M.~Alaburda, Phys. Rev. E {\bf 71}, 051105 (2005). 

\bibitem{TK} J.~Timmer and M.~Konig, Astron. Astrophys. {\bf 300}, 707 (1995). 

\bibitem{fractality}
P.~Meakin, {\em Fractals, Scaling and Growth Far From Equilibrium\/}, Cambridge
  University Press (1998);
E.~Bacry, J.~Delour, and J.~F. Muzy, Phys. Rev. E {\bf 64}, 026103 (2001);
D.R. Bickel, Fractals {\bf 11}, 245 (2003);
J.~W. Lee, K.~E. Lee, and P.~A. Rikvold, arXiv: nlin/0412038  (2004).

\bibitem{white_lorentz}
T.~Lukes, Proc. Phys. Soc. {\bf 78}, 153 (1961);
C.~Heiden, Phys. Rev. {\bf 188}, 319 (1969);
K.~L. Schick and A.~A. Verveen, Nature {\bf 251}, 599 (1974).

\bibitem{point_process}
S.~Thurner, S.~B. Lowen, M.~C. Feurstain et al., Fractals {\bf 5}, 565 (1997);
L.~Telesca, V.~Cuomo, V.~Lapenna, and M.~Macchiato, Fluct. Noise Lett. {\bf
  2}, L357 (2002);
T.~Musha and H.~Higuchi, Jap. J. Appl. Phys. {\bf 15}, 1271 (1976);
A.~J. Field, U.~Harder, and P.~G. Harrison, IEE Proceedings-Communications {\bf
  151}, 355 (2004);
I~Csabai, J. Phys. A {\bf 27}, L417 (1994);
F.~Gr\"{u}neis, Physica A {\bf 123}, 149 (1984);
T.~Musha and K.~Shimizu, Jap. J. Appl. Phys. {\bf 26}, 2022 (1987);
M.~Y. Choi and H.~Y. Lee, Phys. Rev. E {\bf 52}, 5979 (1995);
F.~Gr\"{u}neis, Fluct. Noise Lett. {\bf 1}, R119 (2001);
J.~M. Halley and P.~Inchausti, Fluct. Noise Lett. {\bf 4}, R1 (2004).

\bibitem{KR} B.~Kaulakys and J.~Ruseckas, Phys Rev. E {\bf 70}, 020101 (R) (2004).

\bibitem{power-law}
B.~B. Mandelbrot, {\em Multifractals and 1/f Noise\/}, Springer-Verlag (1999);
B.~B. Mandelbrot, {\em Fractals and Scaling In Finance\/}, Springer-Verlag
  (1997);
R.~N. Mantegna and H.~E. Stanley, {\em An Introduction to Econophysics:
  Correlations and Complexity in Finance\/}, Cambridge University Press (1999);
X.~Gabaix, P.~Gopikrishnan, V.~Plerou, and H.~E. Stanley, Nature {\bf 423}, 267
  (2003).

\bibitem{GK} 
V.~Gontis and B.~Kaulakys, Physica A {\bf 343}, 505 (2004);
V.~Gontis and B.~Kaulakys, Physica A {\bf 344}, 128 (2004).

\end{thebibliography}
\end{document}